\documentclass[10pt,conference,a4paper]{IEEEtran}
\pdfoutput=1
\ifCLASSINFOpdf
   \usepackage[pdftex]{graphicx}
   \DeclareGraphicsExtensions{.pdf,.jpeg,.png}
\else
\fi

\usepackage{fancyhdr} 
\usepackage{amsmath}
\usepackage{algorithmic}
\usepackage{array}
\ifCLASSOPTIONcompsoc
  \usepackage[caption=false,font=normalsize,labelfont=sf,textfont=sf]{subfig}
\else
  \usepackage[caption=false,font=footnotesize]{subfig}
\fi
\usepackage{url}

\usepackage{cases}
\usepackage{xcolor}

\newcommand{\unitspace}{\,}

\begin{document}
\title{Restoration of Images with Wavefront Aberrations}

\author{\IEEEauthorblockN{Claudius Zelenka}
\IEEEauthorblockA{Department of Computer Science\\
Kiel University\\
Germany\\
Email: cze@informatik.uni-kiel.de}
\and
\IEEEauthorblockN{Reinhard Koch}
\IEEEauthorblockA{Department of Computer Science\\
Kiel University\\
Germany\\
Email: rk@mip.informatik.uni-kiel.de}
}

\maketitle

\begin{abstract}This contribution deals with image restoration in optical systems with coherent illumination, which is an important topic in astronomy, coherent microscopy and radar imaging. Such optical systems suffer from wavefront distortions, which are caused by imperfect imaging components and conditions.
Known image restoration algorithms work well for incoherent imaging, they fail in case of coherent images.

In this paper a novel wavefront correction algorithm is presented, which allows image restoration under coherent conditions.
In most coherent imaging systems, especially in astronomy, the wavefront deformation is known. Using this information, the proposed algorithm allows a high quality restoration even in case of severe wavefront distortions.
We present two versions of this algorithm, which are an evolution of the Gerchberg-Saxton and the Hybrid-Input-Output algorithm.
The algorithm is verified on simulated and real microscopic images.
\end{abstract}

\IEEEpeerreviewmaketitle

\newcommand*\diff{\mathop{}\!\mathrm{d}}

\section{Introduction}

\begin{figure}
\centering
\subfloat[Input with synthetic aberrations]{
\includegraphics[width=0.47\columnwidth]{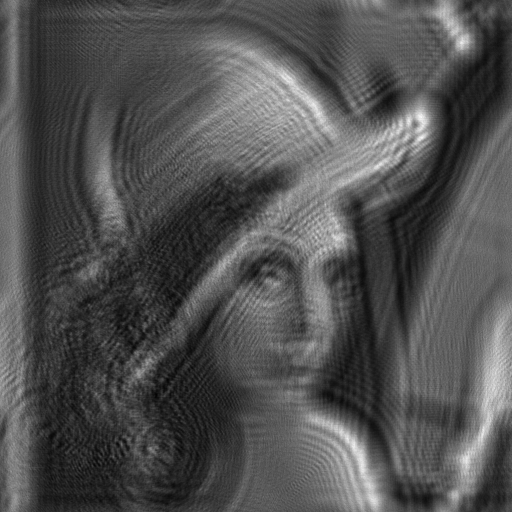}
}
\subfloat[Restored image with 500 WFC-GS iterations ]{
\includegraphics[width=0.47\columnwidth]{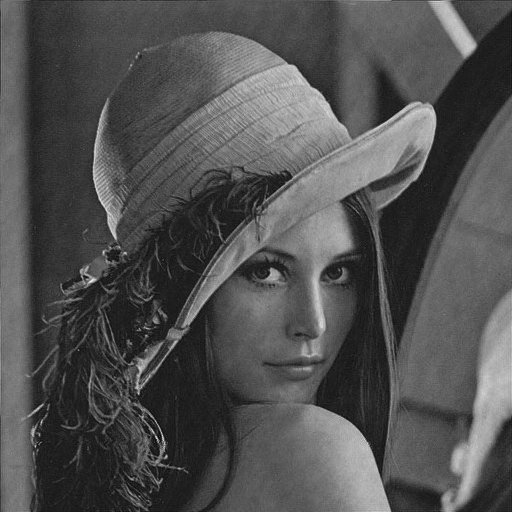}
}
\caption{Input and ouput of the presented restoration algorithm on synthetic blur. Best viewed in digital format.}\label{fig:highlight}
\end{figure}

In this paper an algorithm based on the Gerchberg-Saxton algorithm is presented, which restores images disturbed by optical aberrations in case of coherent or partially coherent illumination. Prior deconvolution algorithm are not capable of this task.  
This algorithm, the wavefront correction algorithm (WFC), can be applied directly on imagery from coherent microscopy and astronomy and deal with the restoration of images blurred by wavefront errors, caused frequently by imperfect optical elements, alignment or by defocus.
An example of blur caused by defocus (a spherical wavefront aberration) and the restored image is shown in Figure \ref{fig:highlight}.
The removal of these aberrations in the context of coherent illumination is important, as coherent imaging conditions occur even in standard microscopic settings (see experimental Section \ref{sec:real_experiments}) and the light from the stars in astronomic imaging fulfills the conditions of spatial coherence. 

Perfect aberration free optics are very costly and hard to design as the initially blurry images of the Hubble space telescope have shown \cite{allen_hubble_1990}. 
The wavefront aberration in diffraction limited, sometimes called "Fourier optics", needs to be less than $\lambda/5$. 
Sometimes a total aberration of even less than $\lambda/10$ is required. 
This is hard to achieve, especially in systems composed of many elements. 

The proposed algorithm can restore high quality images from imperfect coherent optical systems, such as Hubble space telescope before repairs. 
Restoration is possible even if the aberrations are constantly changing, like in earth-bound astronomy, where complex adaptive optics with deformable mirrors are necessary to compensate atmospheric turbulence \cite{hickson_atmospheric_2014}.
A major advantage is that this algorithmic approach is not limited in modulation speed and magnitude in contrast to deformable mirrors.  

Standard image processing deconvolution algorithms such as the Wiener restoration filter \cite{gunturk_image_2012}, the Richardson-Lucy algorithm \cite{lucy_iterative_1974} \cite{richardson_bayesian-based_1972} and modern blind deconvolution algorithms %
\cite{krishnan_fast_2009} are based on the following image formation model, where an image $M$ is formed from the undisturbed image $I$, blur kernel $B$ and additional noise $N$
\begin{equation}
M = I \otimes B  + N.
\end{equation}

This model describes the case of incoherent imaging well.

In case of coherent imaging, as in astronomy or coherent microscopy, the image formation is described by a different model \cite{goodman_introduction_2005}
\begin{equation}
A_m = U \otimes B_a  + N_a\label{eq:coherent},
\end{equation}
where $U$ is the undisturbed complex amplitude and $B_a$ and $N_a$ amplitude blur kernel and noise. Cameras acquire an image by measuring the light intensity distribution over an image sensor or historically a photosensitive film. The magnitude of the amplitude distribution is the square root of the light intensity distribution. $A_m$ is the complex amplitude distribution of the optical wave in the image plane.
The recorded intensity image contains no phase information, therefore the complex part of the amplitude is unknown and standard algorithms cannot be applied, as they cannot deal with interference (see Figure~\ref{fig:non-coherent}). 
Clearly incoherent imaging is linear in intensity, while coherent imaging is linear in amplitude. 
The blur kernels are related \cite{goodman_introduction_2005}:
\begin{equation}
B = (\lvert B_a \rvert) ^2.
\end{equation}

In the following we shall deal with coherent imaging only.
Phase retrieval algorithms solve the task of restoring the unknown phase of a wavefront from measured intensity distributions in Fourier and object plane with additional spatial constraints in the object plane. 
A more detailed overview of phase retrieval algorithms is given in \cite{shechtman_phase_2015} and a performance comparison of various algorithms in \cite{shevkunov_comparison_2014}. Notable recent developments are \cite{netrapalli_phase_2015} and \cite{bian_fourier_2015}. 
In the area of coherent optics phase retrieval algorithms are widely used in many applications from crystallography, holography to ptychography and astronomy. An overview of the applications of the  Gerchberg-Saxton (GS) algorithm and its variations is given in \cite{fienup_phase_2013}.
The classic phase retrieval algorithm is the GS-algorithm \cite{gerchberg_practical_1972}. It is discussed in the following section, together with the Hybrid-Input-Output (HIO) algorithm for phase retrieval by Fienup \cite{fienup_phase_1982}.%

 They form the base of our algorithm, however it is important to note that this paper deals with a different task, namely the correction of wavefront distortions solely from a measured intensity distribution in the image plane.  Additionally some estimate of the wavefront distortion is needed. 
This cannot be accomplished with phase retrieval or the prior non-coherent image restoration algorithms (see Figure \ref{fig:non-coherent}).

Furthermore we use different constraints than prior phase retrieval algorithms, as we introduce an additional virtual focus plane, whose amplitude distribution we require to be real and positive.
Good theoretical convergence properties of the algorithm were confirmed by its application on simulated and real complex images.
A comparison between simulated and real images is performed on a standard bright field microscope with a microscopic lens. The illumination lamp was replaced by a standard LED, which showed satisfactory wavelength coherence. The required spatial coherence could be achieved by adjusting the illumination diaphragm. 

\subsection{The Phase Retrieval Problem} \label{sec:phase_ret}

The amplitude distributions across the real-space object plane $o$ and a Fourier-space plane $A$ are linked by the Fourier transform operator $\mathcal{F}$:
\begin{equation}
o(n) = \mathcal{F}(A)(n) = \int A(k) \exp{(-j 2\pi k n)} \diff k.
\end{equation}

The goal is to recover the phase of $o$ from a set of measurements, while fulfilling the constraints on the object plane and on the Fourier plane.
The choice of constraints by Gerchberg in \cite{gerchberg_practical_1972} is restricting the object magnitude in real- and Fourier-space to the measured magnitudes $|o_0(n)|$ and $|A_0(k)|$.

\begin{figure}
\centering %
\includegraphics[width=0.95\columnwidth,page=1,trim= 4cm 7.5cm 10cm 6cm,clip=true]{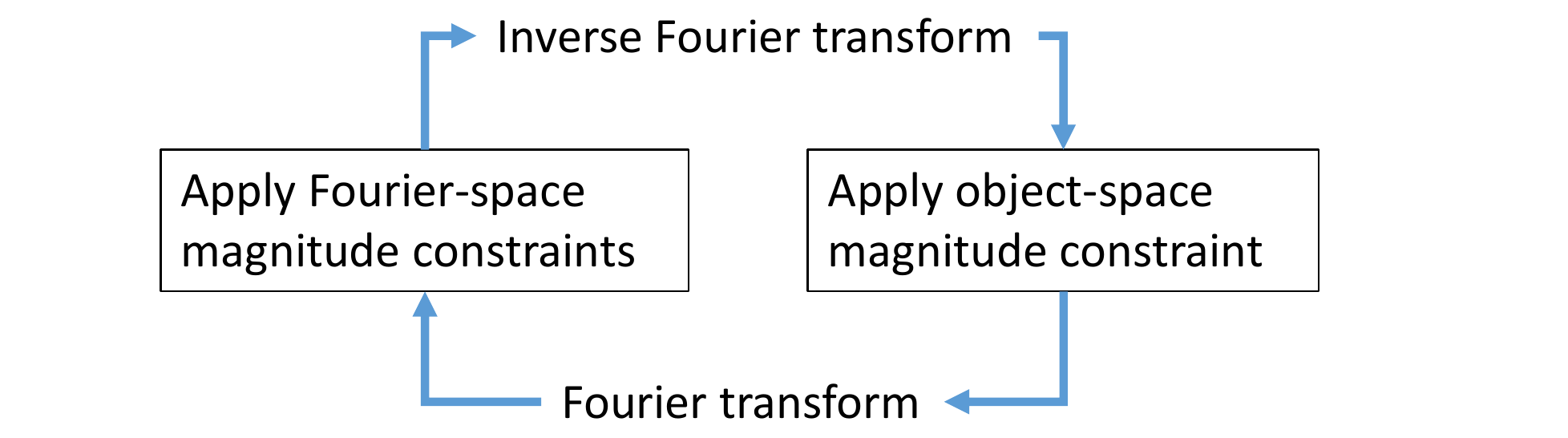}
\caption{The Gerchberg-Saxton algorithm}\label{fig:gs}
\end{figure}

The algorithm starts with an arbitrary phase distribution in $o$ and loops over real-space and Fourier-space, while enforcing the constraints in every iteration $i$, (see Figure \ref{fig:gs}):
\begin{align}
A_i(k) &= \mathcal{F}(o_i)(k)\label{eq:gs1}\\ 
A'_i(k) &= |A_0(k)| \frac{A_i(k)}{|A_i(k)|}\label{eq:fourier_update} \\
o'_i(n) &= \mathcal{F}^{-1}(A'_i)(n)\label{eq:gs3}\\ 
o_{i+1}(n) &= |o_0| \frac{o'_i(n)}{|o'_i(n)|}\label{eq:object_update}
\end{align}

A disadvantage of this basic algorithm is its slow convergence. This was greatly improved by Fienup in \cite{fienup_phase_1982}. His Hybrid-Input-Output (HIO) algorithm  is widely used today.
This algorithm treats the application of Fourier domain constraints as a feedback system with the input $o_i$ in Eq. \ref{eq:gs1} and output $o'_i(n)$  in Eq. \ref{eq:gs3}. Furthermore an additional support constraint is added, which restricts the size of the reconstruction to a prior defined set of points.
The object-space update Eq. \ref{eq:object_update} is changed to:
\begin{equation}
o_{i+1}(n) = \begin{cases}
o'_i(n) & n\in V\\
o_i(n) - \beta o'_i(n)& n \not\in V, \\
\end{cases}
\end{equation}
 where $V$ is a set of points, where the object constraints are valid and $\beta\in \mathrm{R}$ is a parameter influencing the convergence speed. A value commonly used is $0.7$ \cite{fienup_phase_2013}, we follow this convention. For an evaluation of the influence of this parameter see \cite{fienup_phase_1982}.

From an input image, that is violating a constraint, the enforcement of a constraint outputs a result which is the first valid input closest to that violating input. 
This application of a constraint can also be expressed in the framework of projections \cite{youla_image_1982}, where the application of a projection means the enforcement of a constraint. 
Let $A$ be the set of values for which a constraint holds, then we define a projection $P$ of $y$ onto $A$ as the set of values $x \in A$, which are closest to $y$:
\begin{equation}
P_A(y) = \{x \in A, \Vert x-y \Vert = \Vert y-A\Vert \}.
\end{equation}
For practical purposes if the set has more than one value, we choose one of these points. 
For conditions for single-valuedness of the Fienup constraints and a more detailed introduction see \cite{bauschke_phase_2002}.
The application of the constraint in Eq. \ref{eq:object_update} is defined as $P_o$ and the subsequent application of equations \ref{eq:gs1}, \ref{eq:fourier_update} and \ref{eq:gs3} as $P_A$.
The HIO-algorithm now reads:
\begin{equation}
o_{i+1}(n) = \begin{cases}
P_A(o_i)(n) & n\in V\\
o_i(n) - \beta P_A(o_i) (n)& n \not\in V. \\
\end{cases}\label{eq:hio}
\end{equation}

\section{Wavefront Correction Algorithm}
The goal is to develop an image restoration algorithm, based solely on the knowledge of the intensity distribution of the distorted image. %
The light distribution in the focal plane of a thin lens is the Fourier transform of the light distribution of the aperture (Fourier plane). Thus assuming that the input image is captured in perfect focus we can switch between this focal image plane and the Fourier plane by applying the Fourier transform and the inverse Fourier transform, as in the classic phase retrieval algorithms see Section \ref{sec:phase_ret}. 
However, if the image is not captured in perfect focus, this condition no longer holds.  
We need to include the additional wavefront deformation introduced by defocus or other optical aberrations. 

For coherent light incident on a convex thin lens with radius $r$ (assuming the Fresnel approximation \cite{born_principles_1980}) the phase delay is
\begin{equation}
p_d = \exp (j \frac{k}{2 r} (x^2 + y^2)), \label{eq:lens}
\end{equation}
where $k$ is the wave number and $x$ horizontal and $y$ vertical distance from the center of the lens perpendicular to the optical axis. 
 Thus the effect of a thin lens is equivalent to a spherical wavefront deformation, which has the same form as the defocus error. %
 We generally describe these wavefront aberrations using Zernike polynomials \cite{noll_zernike_1976}.

In some applications, as in a typical microscopic setting, we have no sensors to measure the Fourier space magnitude.%
However we know that in an ideal coherent optical system the image of a real object is a real positive function with planar phase.
This is a consequence of the linearity of an aberration free imaging system.

Therefore we can use two constraints: The first constraint is the measured amplitude distribution, which is the square root of the intensity distribution, the second one is the requirement for a real and positive amplitude distribution in the virtual ideal imaging plane, where all aberrations are removed. %
The application of phase aberration function $p_s$ or its inverse $p_s^{-1}$ means a phase shift on the corresponding complex amplitudes. It is defined analogous to the specific defocus phase shift in Eq. \ref{eq:lens}.                  
\subsection{WFC-GS}
We introduce a virtual focus plane $f$, which can be reached from Fourier-space $A$ via the inverse Fourier transform. 
The following equations define the Gerchberg-Saxton (GS) version of the Wavefront Correction Algorithm (WFC-GS) in iteration step $i$, illustrated by Figure~\ref{fig:algo}:

\begin{figure}
\centering
\includegraphics[page=2,width=0.95\columnwidth,trim= 2.7cm 7.5cm 8cm 2cm,clip=true]{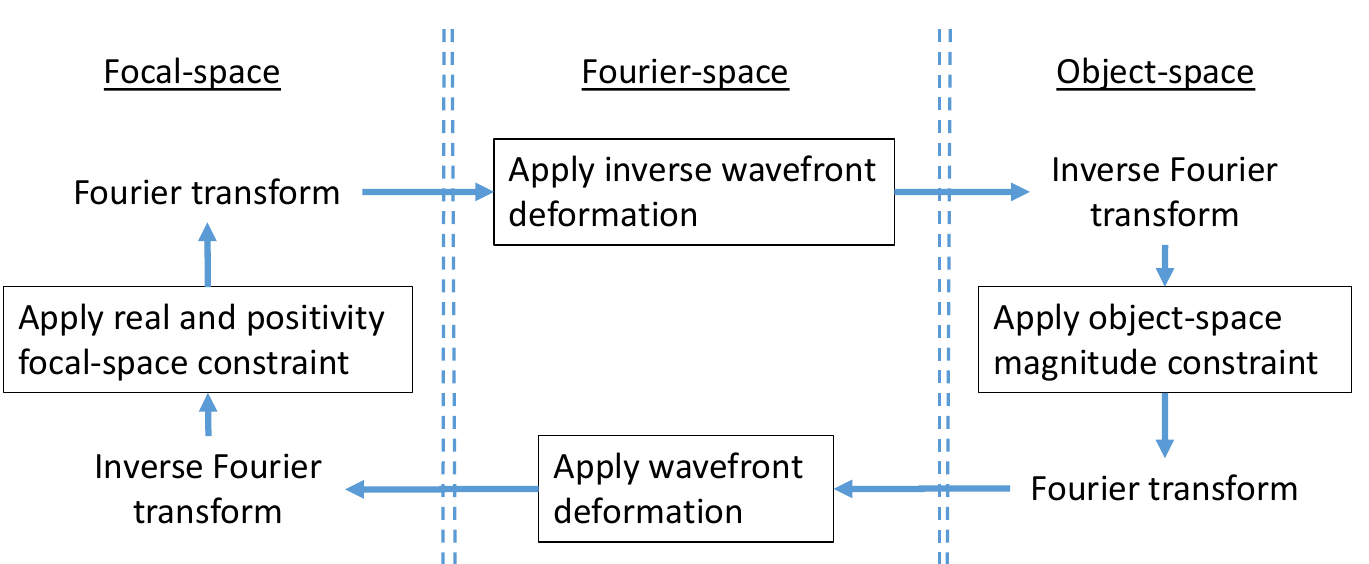}
\caption{Overview of the WFC algorithm}\label{fig:algo}
\end{figure}

\begin{align}
A_i(k) &= \mathcal{F}(o_i)\, p_s (k)  \label{eq:algo1}\\
f_i(n) &= \mathcal{F}^{-1}(A_i)(n)\label{eq:algo2}\\
f'_i(n) &= |f_i(n)|\label{eq:algo3}\\
A'_i(k) &= \mathcal{F}(f'_i)(k) \label{eq:algo4}\\
o_i(n) &= \mathcal{F}^{-1}(A'_i \,p_s^{-1})(n)\label{eq:algo5}\\
o_{i+1}(n) &= |o_0(n)| \frac{o_i(n)}{|o_i(n)|\label{eq:algo6}}.
\end{align}

The algorithm starts similar to the GS-algorithm with the measured $o_0$ as $o_i$. 
In the first step (Equation \ref{eq:algo1}) the Fourier transform is applied. Then the wavefront deformation is applied by multiplying the Fourier space distribution with phase delay $p_s$ of this wavefront. 
Transferred to the virtual focal plane (\ref{eq:algo2}), the real and positive constraint of the distribution is enforced (\ref{eq:algo3}).
Then we transform back into object-space (\ref{eq:algo4}, \ref{eq:algo5}), enforce the object magnitude constraint (\ref{eq:algo6}) and start iteration $i+1$.

The algorithm starts in object space with the measured magnitude distribution.

For the practical application of the algorithm note that the knowledge of the exact wave number of Eq. \ref{eq:lens} is not necessary, as it is only a linear factor with the radius. 
More practical is a measure of maximum phase difference dependent on the image size. 

In \cite{fienup_phase_1982} the weak convergence for the GS-algorithm is shown using an error reduction argument, we expand on this to prove the same property for the GS-version of the wavefront correction algorithm (WFC-GS).

In iteration $i$ the squared error against the real-space plane constraint over the entire image is
\begin{equation}
E_{o,i}^2 = \int_n (o_{i+1}(n) - o'_i(n))^2 \diff n. \label{eq:proof1}
\end{equation}
Similar for the focus plane constraint the squared error is
\begin{equation}
E_{f,i}^2 = \int_n (f_i(n) - f'_i(n))^2 \diff n.\label{eq:proof2}
\end{equation}
The squared error may also be seen as the energy of the error. With Parseval's theorem it follows that
\begin{equation}
E_{f,i}^2 = \int_k (A_i(k) \, p_s^{-1} - A'_i(k) \,p_s^{-1})^2 \diff k.\label{eq:proof3}
\end{equation}
Because of the energy conservation of optical light transport, we obtain
\begin{equation}
E_{f,i}^2= \int_k (A_i(k) - A'_i(k))^2 \diff k.\label{eq:proof4}
\end{equation}
This allows a second application of Parseval's theorem:
\begin{equation}
E_{f,i}^2 = \int_n (o_i(n) - o'_i(n))^2 \diff n. \label{eq:proof5}
\end{equation}
Because for any $n$
\begin{equation}
 (o_i(n) - o'_i(n))^2  \leq  (o_{i+1}(n) - o'_i(n))^2,\label{eq:proof6}
\end{equation}
it follows with equations \ref{eq:proof1} and \ref{eq:proof5} that
\begin{equation}
E_{o,i}^2  \leq E_{f,i}^2. \label{eq:proof7}
\end{equation}
With the same arguments for $f$, clearly
\begin{equation}
(f_{i+1}(n) - f'_{i+1}(n))^2 \leq (f_i(n) - f'_i(n))^2. \label{eq:proof8}
\end{equation}
From this it follows with equations \ref{eq:proof2} and \ref{eq:proof5} that
\begin{equation}
E_{f,i+1}^2 \leq E_{0,i}^2. \label{eq:proof9}
\end{equation}
Combined with equations \ref{eq:proof7} we see
\begin{equation}
E_{f,i+1}^2 \leq E_{0,i}^2\leq E_{f,i}^2\\
\end{equation}
and
\begin{equation}
E_{f,i+1}^2 \leq E_{f,i}^2.
\end{equation}
Therefore the error decreases or stagnates with every iteration. 
This result shows the weak convergence, but does not guarantee that the algorithm always converges to a global minimum. 

\subsection{WFC-HIO}

As a further step we adapt the HIO-algorithm to focal-space constraints and wavefront correction. 
We denote the application of Eq. \ref{eq:algo6} as $P_o$ and the application of Eq. \ref{eq:algo1} to \ref{eq:algo5} as $P_f$. 
Because we lack information about the object support, we cannot directly apply the standard HIO-definition Eq. \ref{eq:hio}. Let $\boldsymbol{1}_V$ be the characteristic function of the set of points $V$ where the constraints (such as object support) are valid  and $\boldsymbol{1}_{\boldsymbol{C}V}$ of its complement.
With \cite{bauschke_phase_2002} we write the HIO-version of the wavefront correction algorithm (WFC-HIO) as:

\begin{align}
o_{i+1}(n) &= \boldsymbol{1}_V P_f(o_i)(n) + \boldsymbol{1}_{\boldsymbol{C}V}(o_i - \beta P_f(o_i))(n)\\
&= \boldsymbol{1}_V P_f(o_i)(n) + (\boldsymbol{1} - \boldsymbol{1}_{V}(o_i - \beta P_f(o_i))(n)\\
&=\boldsymbol{1}_V((1 + \beta) P_f(o_i) - o_i) + (I - P_f)(o_i)(n)\\
&= (P_o((1 + \beta) P_f - I) + I - P_f)(o_i)(n)
\end{align} 

and can apply it on images with wavefront aberration in the following section, without requiring object support information.

\section{Experimental Results} \label{sec:experiments}
To verify the algorithm both in correctness of the results and applicability, we apply it both on synthetic images with simulated aberrations and to real microscopic images.
\subsection{Evaluation on Synthetic Images} \label{sec:simu_experiments}

\begin{figure}
	\centering
\subfloat[Original image]{
\includegraphics[width=0.42\columnwidth]{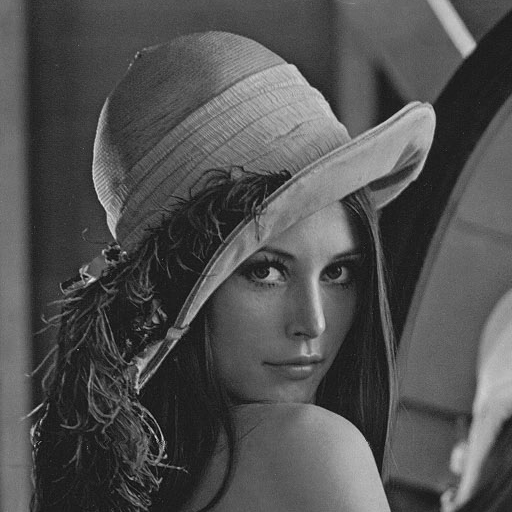}\label{fig:lena_original}
}~
\subfloat[Distorted by wavefront aberration in Figure \ref{fig:arbiWavefront}]{
\includegraphics[width=0.42\columnwidth]{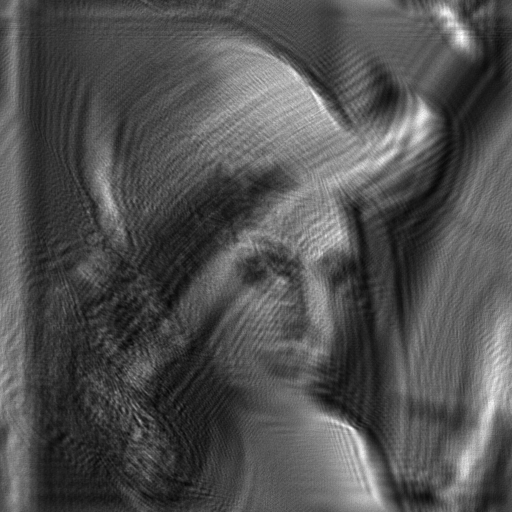}\label{fig:distort_arbiwavefront}
}
\caption{Synthetic image with arbitrary wavefront aberration. }\label{fig:restore_arbiwavefront}
\end{figure}

\begin{figure}
\centering
\subfloat[Defocus, spherical aberration]{
\includegraphics[width=0.38\columnwidth,trim= 0cm 0cm 17cm 25.5cm,clip=true]{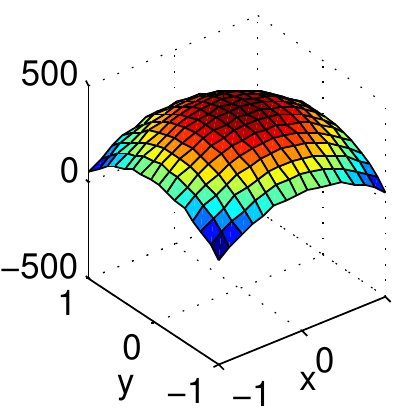}\label{fig:defocusWavefront}
}
\subfloat[A combination of defocus, astigmatism and coma]{
\includegraphics[width=0.38\columnwidth,trim= 0cm 0cm 17cm 25.5cm,clip=true]{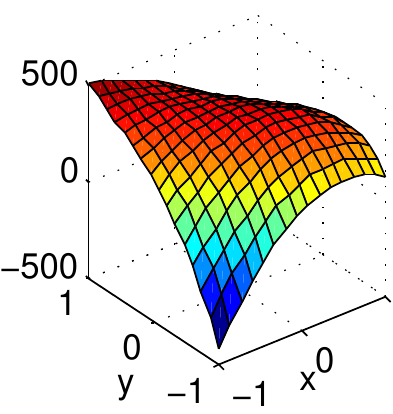}\label{fig:arbiWavefront}
}
\caption{Wavefront aberrations as phase shift over an aperture plane.}\label{fig:wavefronts}
\end{figure}

For the visual verification of restoration quality, we choose a test image with sufficiently high contrast and fine structure (see Figure \ref{fig:lena_original}).
As a first test synthetic defocus aberration is applied on the image by convolving it with the PSF generated from the spherical wavefront deformation in Figure \ref{fig:defocusWavefront}.
The task of the algorithm is now to restore the undisturbed image from this aberrated image.  
We initialize the unknown phase in real-space with zero and apply the WFC-GS algorithm using the known value of the optical aberration.
The synthetic image together with the very convincing restoration is depicted in Figure \ref{fig:highlight}. %
The results in Figure \ref{fig:highlight} show a near perfect restoration even for this very strong aberration and thereby verify that the algorithm works as theoretically predicted.\\

\begin{figure}
\centering
\subfloat[Richardson-Lucy algorithm \cite{richardson_bayesian-based_1972} \cite{lucy_iterative_1974}, implementation see \cite{matlab_version_2015}  ]{
\includegraphics[width=0.42\columnwidth]{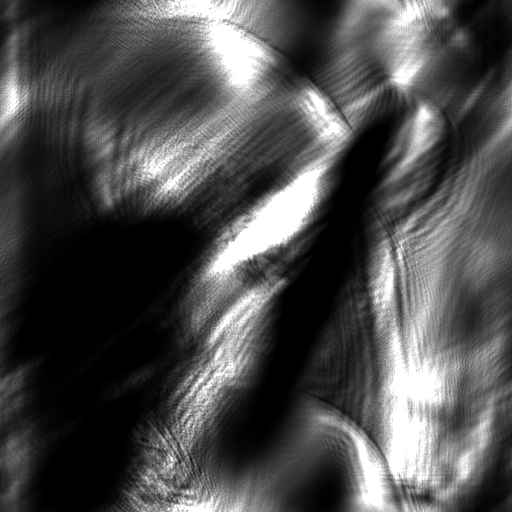}\label{fig:richardson}
}~
\subfloat[Algorithm of Krishnan and Fergus \cite{krishnan_fast_2009} \protect\footnotemark]{
\includegraphics[width=0.42\columnwidth]{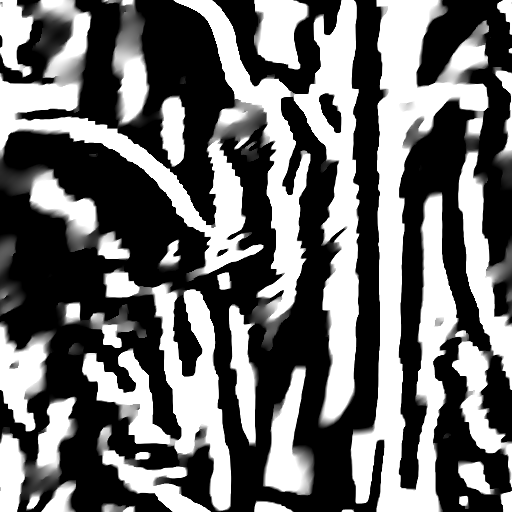}\label{fig:fergus}
}
\caption{Restoration results of Figure \ref{fig:distort_arbiwavefront} with standard deconvolution algortihms, PSF generated from aberration in Figure \ref{fig:arbiWavefront}. }\label{fig:non-coherent}
\end{figure}

\begin{figure}
\centering
\subfloat[Restoration with 500 WFC-HIO iterations, PSNR=$41.5 \unitspace\text{dB}$]{
\includegraphics[width=0.40\columnwidth]{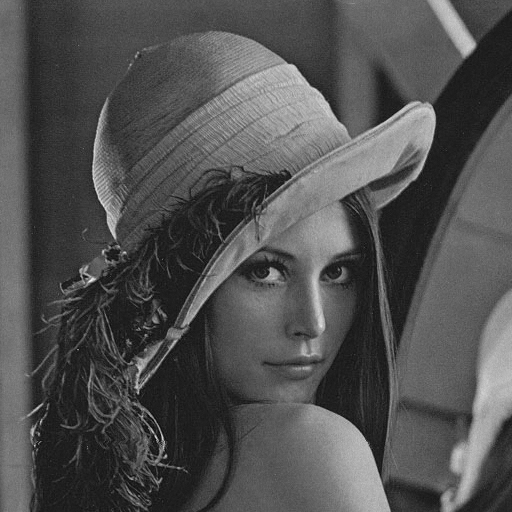}\label{fig:highiteration}
}~
\subfloat[Restoration with 10 WFC-HIO iterations, PSNR=$33.2\unitspace \text{dB}$]{
\includegraphics[width=0.40\columnwidth]{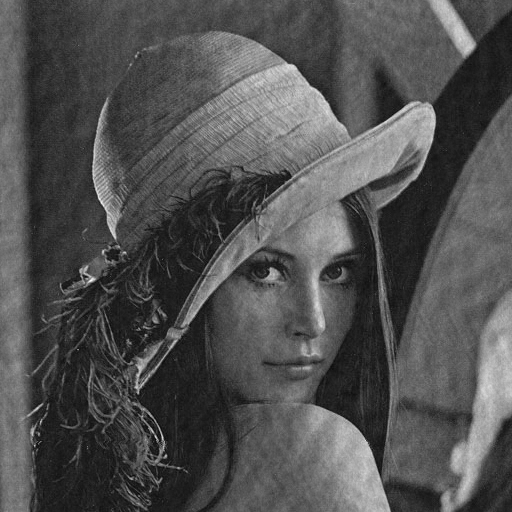}\label{fig:lowiteration}
}

\caption{Comparison between the restored image after 500 iteration and after 10 iterations. The aberrated input image is shown in Figure \ref{fig:distort_arbiwavefront} and the reference image in Figure \ref{fig:lena_original}. The images are best viewed in digital format.}\label{fig:simulated}
\end{figure}

\footnotetext{implementation from \url{http://cs.nyu.edu/~dilip/research/fast-deconvolution/}}
Above we have shown that the GS-version of wavefront correction image restoration algorithm (WFC-GS) converges in the weak sense.
For the Gerchberg-Saxton algorithm for phase retrieval quite high iteration counts of over $2000$ have been reported to overcome stagnation in the results \cite{fienup_phase_1982}, which makes the GS-algorithm unpractical and was the primary reason for the improvements developed by Fienup. 
To test the convergence of the presented algorithm, we measure the peak signal to noise ratio (PSNR) between the focal image in every iteration and the ideal original. 

To verify that the algorithm is not limited to defocus, we apply a combination of synthetic spherical aberration, astigmatism and coma original image, by convolving it with the PSF generated from the Zernike polynomials of the aberration \cite{noll_zernike_1976}. We start with this strongly distorted image in Figure \ref{fig:distort_arbiwavefront}, where the wavefront aberration is depicted in Figure \ref{fig:arbiWavefront}. 
Very high PSNR values of over $40\unitspace \text{dB}$ prove that the result of the algorithm in Figure \ref{fig:highiteration} is not only visually convincing, but also leads to correct results, even in case of complex wavefront aberrations. This is better than the PSNR values of $35 \unitspace \text{dB}$  reported by \cite{welstead_fractal_1999} for high quality compression algorithms.

\begin{figure}
\centering
\includegraphics[width=0.70\columnwidth,trim= 0cm 0cm 13cm 22cm,clip=true]{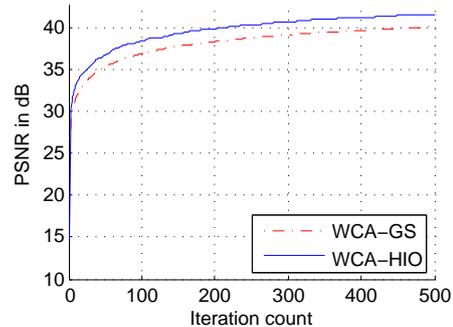}

\caption{Plot of PSNR per iteration for synthetic aberration, input image in Figure \ref{fig:distort_arbiwavefront}.}\label{fig:convergence}
\end{figure}

Comparing both algorithms Figure \ref{fig:convergence}, note the logarithmic scale, we see that the WFC-HIO shows the expected superior convergence. The WFC-HIO reaches a PSNR of $38.3\unitspace\text{dB}$ in iteration 100, while the GS-version requires twice the number of iterations for this value. The difference in iteration $200$ is $1.5\unitspace \text{dB}$. Overall both algorithms show good performance and results. 

The curves show a sub-linear rate of convergence and that many iterations are necessary for near-perfect results, however even after only 10 iterations the WFC-HIO shows good results, with a PSNR of $33.2 \unitspace\text{dB}$, see Figure \ref{fig:lowiteration}. For a $512\times512$ pixel image, on an $4\unitspace \text{GHz}$ desktop processor the time spent per iteration is $14\unitspace \text{ms}$, dominated by the fast Fourier transform.

The results of the Richardson-Lucy algorithm and a modern deconvolution algorithm by Krishnan and Fergus \cite{krishnan_fast_2009}  in Figure \ref{fig:fergus} confirm that on images with such strong aberrations incoherent deconvolution algorithms are not applicable.

\subsection{Evaluation on Real Images}\label{sec:real_experiments}

For the evaluation on real data, we choose a standard transmissive bright field microscopic setting.
A standard USAF 1951 target is back-illuminated with a $650\unitspace\text{nm}$ LED. %
The transmitted light is captured by a standard microscopic lens, the eyepiece was replaced by a digital camera. %
Before the application of the algorithm, the gamma curve of the images is linearized and edge tapering to reduce ringing artifacts is applied. 
We use the Matlab edge tapering function \cite{matlab_version_2015}, the same edge tapering technique as in \cite{fergus_removing_2006}. 
Also the pixel offset and scale need to be adjusted.

\begin{figure}
\centering
\subfloat[Measured image at focus]{
\centering
\includegraphics[width=0.42\columnwidth]{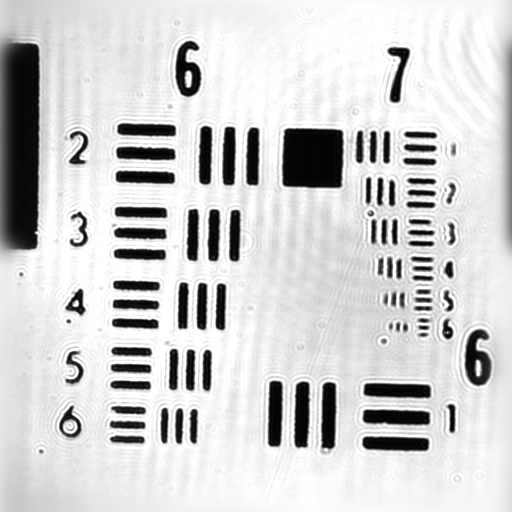}\label{fig:realsharp}}~
\subfloat[Synthetic defocus on (a), same strength of defocus used as for the restoration of image (c)]{
\includegraphics[width=0.42\columnwidth]{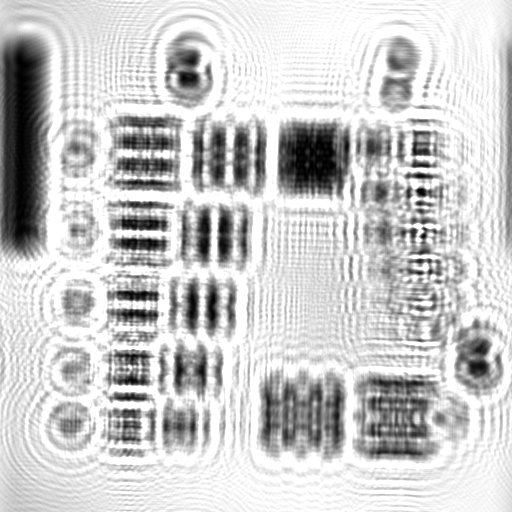}\label{fig:simudefocus}}
\label{fig:verify}
\centering
\subfloat[Measured intensity distribution at $1.95\unitspace\text{mm}$ from sharp focus image (a)]{
\centering\includegraphics[width=0.42\columnwidth]{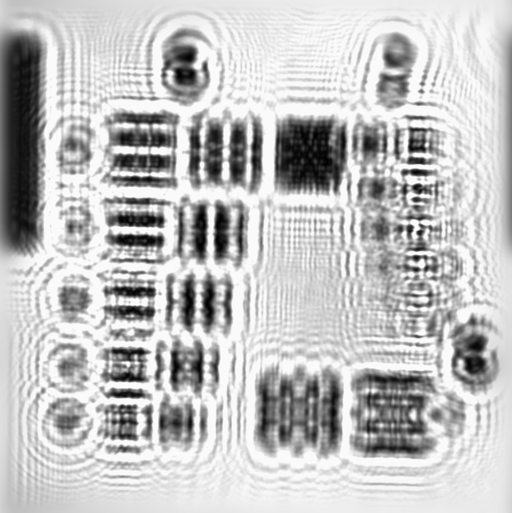}\label{fig:realdefocus}}~
\subfloat[Restored image from (c) with $75$ WFC-GS iterations]{
\centering
\includegraphics[width=0.42\columnwidth]{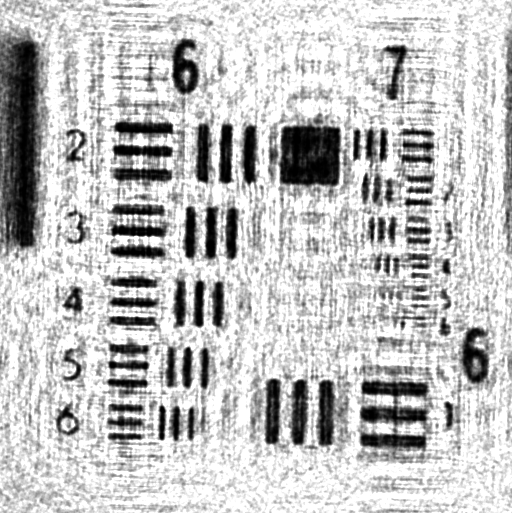}\label{fig:realrestoration}}
\caption{Comparison of sharp original, measured defocused image, synthetic defocus image and restored image. }\label{fig:real}
\end{figure}

For this imaging setup we acquired a series of images with $50\unitspace\mu\text{m}$ steps of the sharpest image on the optical axis. In Figure \ref{fig:realsharp} the sharp image of the target is shown, in Figure \ref{fig:realdefocus} the image at a distance of $1950\unitspace\mu\text{m}$ from the focal plane. 
We choose such a strong defocus, to demonstrate the effectiveness of the algorithm.

The restoration is depicted in Figure \ref{fig:realrestoration}. The algorithm achieves very good results even for this very strong wavefront aberration. None of the numbers are identifiable in the input image, but good contrast and high readability in the restoration is achieved. 
We consider acquisition noise and non-perfect coherence of the LED illumination as the source of the ringing artifacts.

Although an LED is far less coherent than a laser, our simple experiment shows that it can be used for a coherent illumination in microscopic applications.
For a detailed discussion on spatial and frequency coherence see \cite{goodman_introduction_2005}.

A synthetic defocus image with the same defocus parameters as used for the restoration is shown in Figure \ref{fig:simudefocus}. 
This image is at first glance almost identical to the real measurement in Figure \ref{fig:realdefocus}, which again verifies the correctness of the measurement.

\section{Conclusion}

The presented algorithm allows the restoration of coherent images even in case of strong wavefront aberrations.
This should permit an essential cost reduction in coherent optical systems, for example in astronomy and coherent microscopy. Superb images can be achieved even with low quality optical components.

In most applications the wavefront error of the optical system is known, it can be measured using an interferometer or a point light source. In astronomy, it is common to use an artificial laser guide star to measure the wavefront aberrations induced by the turbulence of the atmosphere \cite{hickson_atmospheric_2014}. Adaptive optics, such as deformable mirrors are then used to compensate these wavefront deformations. With the new algorithm the deformable mirrors and their limitations can be avoided. Microscopic applications of adaptive optics with deformable mirrors \cite{booth_adaptive_2014} should also benefit from these improvements.
Moreover, the WFC-algorithm can improve the performance of optical systems, in all cases where the implementation of a deformable mirror is not feasible. 

The high iteration speed, together with the restoration result with $10$ iterations of the WFC-HIO (Figure \ref{fig:lowiteration}) demonstrate that if decent, yet non-perfect restorations are acceptable, real-time application of this algorithm is possible. 
This speed already allows parameter optimization and fine-tuning of the wavefront aberration by measuring the image sharpness.
There are many possibilities for a faster implementation of this algorithm, e.g. on a GPU.
By changing the value of the spherical aberration, we should be able to simulate autofocus.
Furthermore, scanning through different values of the spherical aberration could be an interesting feature in the evaluation of coherent microscopic samples. 
An obvious choice as general optimization parameters would be the coefficients of the Zernike polynomials.

Our future work will be focused on further development, optimization and practical implementation of the algorithm.
\section*{Acknowledgment}
This work has partly been supported by the German Research Foundation (DFG) Cluster of Excellence FUTURE OCEAN under proposals CP1331 and CP1525.

\bibliographystyle{IEEEtran}
\bibliography{IEEEabrv,cleanbib}
\end{document}